\documentclass[12pt,preprint]{aastex}
\usepackage{emulateapj5}
\begin {document}

\title{High Spatial Resolution Mid-IR Imaging of V838 Monocerotis: Evidence of New Circumstellar Dust Creation}

\author{John P. Wisniewski\altaffilmark{1}, Mark Clampin \altaffilmark{1}, 
Karen S. Bjorkman\altaffilmark{2}, Richard K. Barry\altaffilmark{1}}

\altaffiltext{1}{NASA GSFC Exoplanets and Stellar Astrophysics Lab Code 667, 
Greenbelt, MD 20771, John.P.Wisniewski@gmail.com, Mark.Clampin@nasa.gov, Richard.K.Barry@nasa.gov}
\altaffiltext{2}{Ritter Observatory, MS \#113, Department of Physics
and Astronomy, University of Toledo, Toledo, OH 43606, Karen.Bjorkman@utoledo.edu}

\begin{abstract}

We report high spatial resolution 11.2 and 18.1 $\mu$m imaging of V838 Monocerotis obtained 
with Gemini Observatory's Michelle instrument in 2007 March.  Strong emission is observed from the unresolved 
stellar core of V838 Mon in our Gemini imagery, and is confirmed by Spitzer MIPS 24 $\mu$m imaging 
obtained in 2007 April.  The 2007 flux density of the unresolved 
mid-infrared emission component is $\sim$2 times brighter than that observed in 2004.  No clear change 
in the net amount of 24$\mu$m extended emission is observed between the 2004 and 2007 epoch Spitzer 
imagery. 
We interpret these data as evidence that V838 Mon has experienced a new circumstellar dust creation event.  
We suggest that this newly created dust has condensed from the expanding ejecta produced from V838 Mon's 
2002 outburst events, and is most likely clumpy.  We speculate that one (or more) of these 
clumps might have passed through the line-of-sight in late 2006, producing the brief multi-wavelength photometric 
event reported by \citet{bon06} and \citet{mu07b}.  We detect no evidence of extended emission above a level of 
$\sim$1 mJy at 11.2 $\mu$m and $\sim$7 mJy at 18.1 $\mu$m over radial distances of 1860 - 93000 AU 
(0$\farcs$3-15$\farcs$0) from the central source.  Using the simple assumption that ejecta material expands at 
a constant velocity of 300-500 km s$^{-1}$, this gap of thermal emission suggests that no significant prior 
circumstellar dust production events have occurred within the past $\sim$900-1500 years. 

\end{abstract}


\section{Introduction} \label{introduction}

V838 Monocerotis experienced three photometric outbursts in 
early 2002, beginning with an initial event on 2002 January 6.6 \citep{bro02} and continuing with 
events in 2002 February and 2002 March \citep{kim02,mun05}.  The 
star's spectrum cooled through the F, G, and K spectral types throughout 
February \citep{mun07}, transitioned to a late M-type giant by 2002 April 16 \citep{rau02}, and had 
cooled to a L-type supergiant appearance by 2002 October \citep{eva03}.  The circumstellar ejecta from the 
outburst events was not spherically symmetric, as revealed by the intrinsic polarization detected until 
2002 February 13 \citep{wi03a,des04}; \citet{wis07} estimated that  
ejecta located interior to the region producing H$\alpha$ and Ca II emission was flattened by a 
\textit{minimum} of 10\% from a spheroidal shell.  A renewed intrinsic 
polarization component observed on 2002 October 22 might be attributable to 
a new source of asymmetrical scatterers in V838 Mon's circumstellar envelope or to a change in the  
opacity of the existing envelope \citep{wi03b}.

V838 Mon produced a spectacular light echo first detected on 2002 February 17 
by \citet{hen02}, which evolved dramatically over time \citep{bon03}.  
\citet{ban06} imaged V838 Mon with the \textit{Spitzer Space Telescope} 
at 24, 70, and 160$\mu$m, and attributed the observed excess of thermal infrared (IR) emission from the 
unresolved central source as evidence of the formation of circumstellar dust produced by the 2002 
outbursts.   An IR excess was also observed from the spatially resolved light echo, and \citet{ban06} 
concluded that the source of this emitting dust could have both interstellar and circumstellar components.

V838 Mon resides in a young open cluster \citep{afs07} at a distance of 
6.2 $\pm$1.2 kpc \citep{spa08}.  It has a known, albeit unresolved, B3V binary companion 
\citep{mun05}.  The nature of V838 Mon prior to outburst is a subject of active 
debate (c.f. \citealt{mun05, tyl05}); a wide variety of mechanisms 
have been proposed to explain the eruptive events in V838 Mon, including 
a shell thermonuclear event \citep{mun05}, a born-again star which experienced an accretion 
event \citep{law05}, a stellar merger \citep{tyl06}, and a star which engulfed several planets 
\citep{ret06}.  Determining the amount and spatial distribution of dust within the V838 Mon line of sight, and 
quantifying how much is circumstellar versus interstellar in nature, would help constrain whether its outburst 
mechanism was a variant of a repeatable phenomenon, like a thermonuclear 
runaway or nova-like explosion, or more likely a singular event, like a stellar merger.

\section{Observations} \label{obs}

We obtained N$^{'}$ (11.2$\mu$m) and Q$_{a}$ (18.1$\mu$m) imaging of V838 Mon on 
2007 March 21-22 using Gemini Obervatory's Michelle imager.  Michelle is a 320 x 240 pixel array 
which has a 32$\farcs$0 x 24$\farcs$0 field of view (0$\farcs$1 per pixel).  
All observations were obtained using a standard chop-nod mode with the same instrument configuration: a 
detector position angle of 110$^{\circ}$ east of north, a chop angle of 130$^{\circ}$ east of north, 
and the maximum chop throw of 15$\farcs$0.  The  weather and seeing conditions during our observing 
runs were stable, 
with FWHM values of individual chop-nod pairs deviating from the average values quoted in 
Table \ref{obstable} by less than 0$\farcs02$ on 2007 March 21, 0$\farcs$05 on 2007 March 22, 
0$\farcs$04 on 2007 August 6, and 0$\farcs$01 on 2007 August 9.  

\begin{figure}
\includegraphics[scale=0.25]{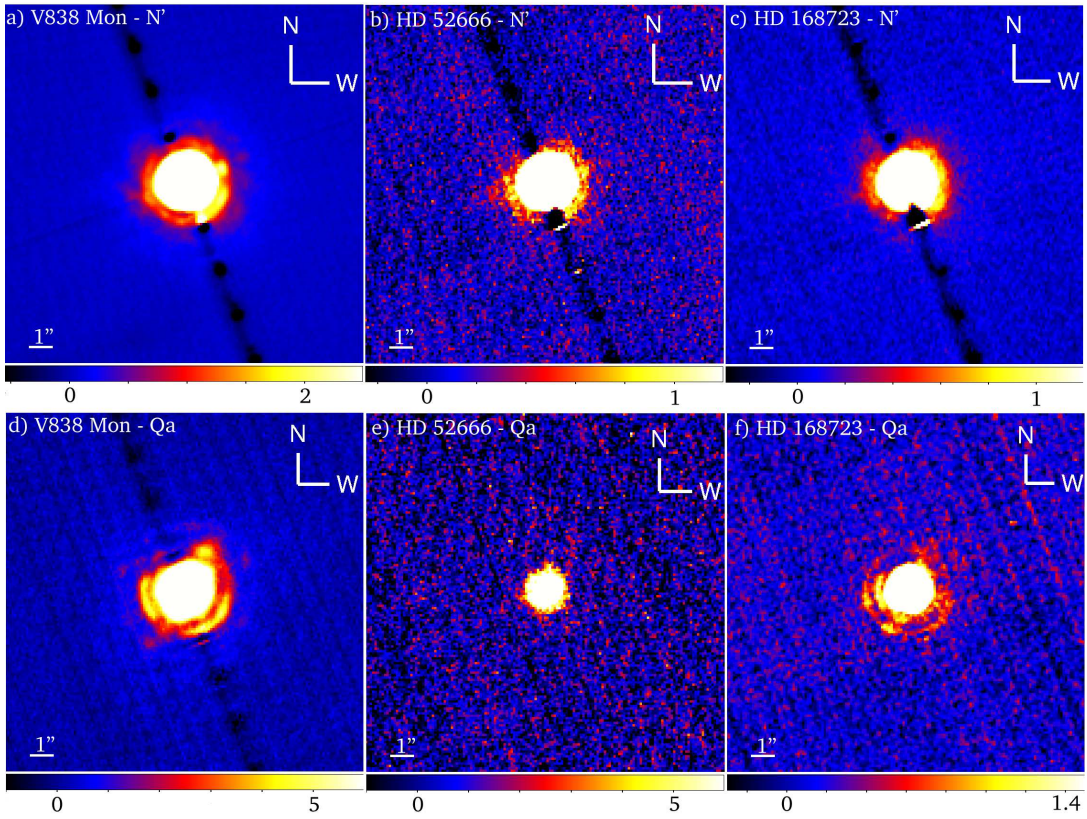} 
\figcaption[f1.eps]{Reduced imagery spanning a 15$\farcs$0 x 15$\farcs$0 field of view and plotted on a linear 
scale in units of mJy.  The displayed panels include (a) N$^{'}$-band V838 Mon; (b) N$^{'}$-band HD 52666 
(PSF-star); (c) N$^{'}$-band HD 168723 (deep PSF-star); (d) Qa-band V838 Mon; (e) Qa-band HD 52666 
(PSF-star); and (f) Qa-band HD 168723 (deep PSF-star) imagery.  The diagonal dark lanes seen prominently in 
panels a-d are known detector artifacts which arise when Gemini's Michelle observes bright sources. \label{rawfig}}
\end{figure}

The data, summarized in Table 1, were reduced with procedures similar to those prescribed in Gemini's \textit{midir} package.  Following 
combination of individual chop-nod pairs, known image artifacts such as vertical and horizontal ``staircases'' 
were characterized by sampling 15 pixel-wide regions along column and row borders, and removed 
via custom IDL routines.  Cleaned images were registered to a common position and examined 
for evidence of anomalous elongation along the chop direction, which is known to characterize a 
fraction of all Michelle data since early 2007.  Our final combined images were flux calibrated using 
observations of Cohen standards \citep{coh99} HD 66141 and HD 60522 
obtained immediately following our V838 Mon observations.  Figure \ref{rawfig} provides a 15$\farcs$0 x 
15$\farcs$0 view of our reduced V838 Mon and PSF-star data, plotted on a linear scale in units of mJy.

V838 Mon was also imaged with the Multiband Imaging Photometer for Spitzer (MIPS; \citealt{rie04}) on 
2007 April 10 by program 30472 (PI K. Su).  These data were reduced using version 16.1.0 of the Spitzer pipeline.  
Earlier epoch MIPS imaging from 2004 October 14 (24$\mu$m) and 2005 April 2 (70$\mu$m) have been 
previously presented in \citet{ban06}.

\section{Analysis} \label{results}
\subsection{Photometry of the Unresolved Point Source} \label{phot}

We used simple aperture photometry to extract flux densities for the unresolved V838 Mon point source in our 
Gemini data.  We explored the use of 5 circular apertures having radii ranging in size from 0$\farcs$6-3$\farcs$0 for our 
N$^{'}$ data and from 0$\farcs$7-3$\farcs$0 for our Q$_{a}$ data, and found consistent results from each.  
The mean flux density values from these apertures, 29.54$\pm$0.14 Jy at 11.2$\mu$m and 36.43$\pm$0.82 Jy 
at 18.1$\mu$m, are cited in Table \ref{obstable}.  The quoted errors are the  
standard deviations from our different apertures, and provide an estimate of the internal uncertainty in our 
measurements ($\sim$1\%).  Additional uncertainties in the absolute flux calibration, owing to the internal 
accuracy of the Cohen flux standards and changes in the atmospheric conditions between our observations, 
are expected to be of order several percent for bright sources observed in stable conditions (K. Volk 2008, 
personal communication).

Spitzer's MIPS observed V838 Mon in 2007, $\sim$2 weeks after our Gemini data were obtained.  The 24$\mu$m 
imagery was heavily saturated at the core.  \citet{ban06} reported similar saturation effects in their 2004 24$\mu$m 
MIPS data.  These authors fit a scaled PSF-star to match the brightness of 
V838 Mon's first airy ring, and replaced the saturated V838 Mon core by the core of this scaled PSF-star to recover 
a flux density measurement of the unresolved V838 Mon central source (Table \ref{obstable}).  
To extract similar measurements for the 2007 24$\mu$m MIPS data, we simply scaled and registered the 
2004 and 2007 V838 Mon data to match their unsaturated PSF wing regions.  The optimal scaling factor 
yielded a 24$\mu$m flux density of 29.69$\pm$ $^{1.95}_{1.91}$ Jy (Table \ref{obstable}).  Note that the quoted 
error represents the uncertainty we estimated for our image scaling ($\sim$1.5\%) along with the 5\% absolute flux 
uncertainty of the 2004 Spitzer flux densities reported by \citet{ban06}.  Using simple aperture 
photometry, we also 
extracted a flux density for the 2007 April Spitzer MIPS 70$\mu$m observation of 7.34$\pm$0.73 Jy 
(Table \ref{obstable}).

The 2007 March Gemini 18.1$\mu$m flux density is consistent with the 2007 April Spitzer 24$\mu$m flux 
density at the 3$\sigma$ level; these flux densities are a factor of two stronger than that observed 
in 2004 October observations.  Similarly, the 2007 April Spitzer 70$\mu$m flux density is $\sim$2 times stronger than
that observed in 2005 April.  The amount of excess IR emission associated with the unresolved central source 
of V838 Mon has clearly increased between 2004/2005 and 2007. 

\subsection{Extended Emission} \label{ee}

Gemini's Michelle provides a 10-fold improvement in spatial resolution compared to Spitzer's MIPS; hence, our data 
provide us the unique opportunity to search for mid-IR extended emission interior to that probed by \citet{ban06}.  
Comparison of median azimuthally averaged radial profiles of our 11.2$\mu$m and 18.1$\mu$m imagery of 
V838 Mon with a similar color PSF-star observation (HD 52666; M2 III), obtained immediately before our V838 Mon 
data, reveal no convincing evidence of extended emission nearby the stellar core.  Measured median FWHM values 
of V838 Mon and 
our PSF-star are the same at 11.2 and 18.1 $\mu$m (Table \ref{obstable}) to within the maximum 
observed deviation amongst chop-nod pairs ($<$0$\farcs$02 at 11.2 $\mu$m and $<$0$\farcs$05 at 
18.1 $\mu$m), which quantitatively supports this conclusion.

\begin{figure}
\includegraphics[scale=0.3]{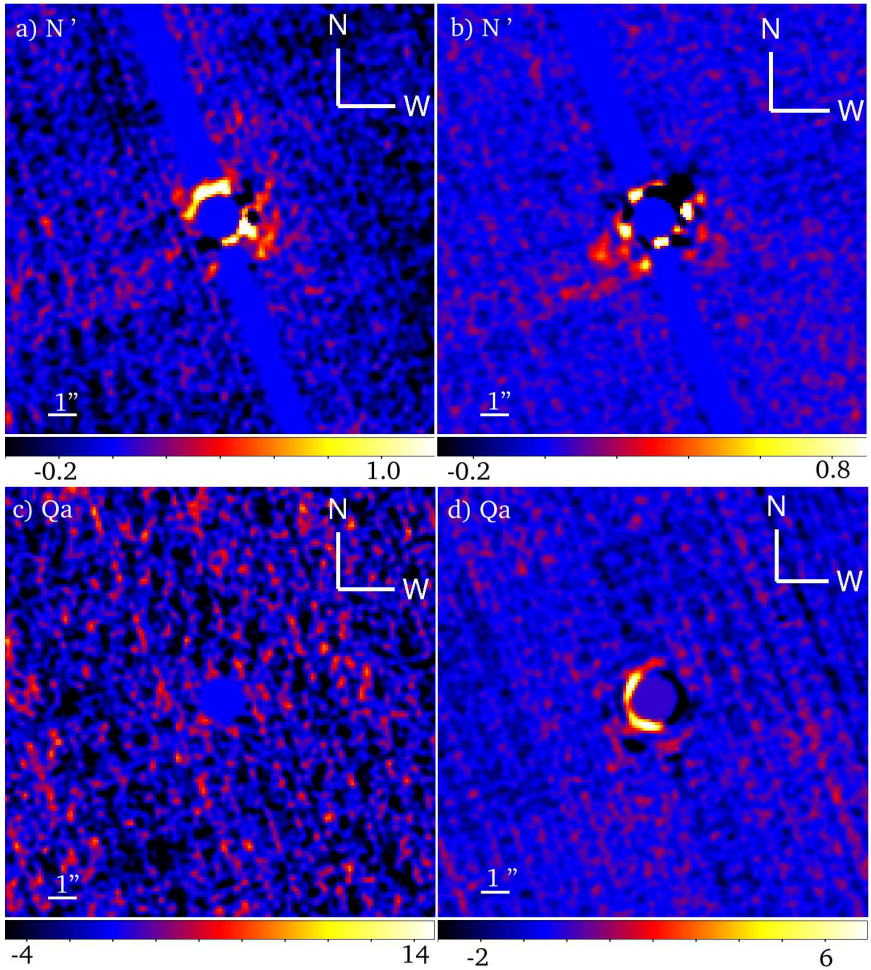} 
\figcaption[f3.eps]{PSF-subtracted V838 Mon imagery plotted on different linear scales in units of mJy.  The field of 
view, 15$\farcs$0 x 15$\farcs$0, lies entirely inside of region of extended emission detected by Spitzer's 
MIPS in 2004 \citep{ban06}.  The displayed panels include (a) N$^{'}$-band imagery using 
HD 52666 as the PSF-star; (b) N$^{'}$-band imagery using HD 168723 as the PSF-star; (c) Qa-band 
imagery using HD 52666 as the PSF-star; and (d) Qa-band imagery using HD 168723 as the PSF-star.  Regions 
inside a radial distance of 0$\farcs$8 (8-pixels) were contaminated by PSF subtraction residuals and are 
therefore masked in the displayed imagery.  We detect no clear evidence of excess extended emission in 
these data. \label{psffig}}
\end{figure}

To search for extended emission outside of the immediate vicinity of the stellar core, we used both 
contemporaneously observed PSF-star observations (HD 52666; Table \ref{obstable}) 
and longer integrations of a PSF-star obtained several months after our V838 Mon observations (HD 168723; 
Table \ref{obstable}) to model and subtract the PSF from our V838 Mon imagery.  The deeper PSF-star 
observations exhibit marginally different FWHM values as compared to V838 Mon, likely attributable to minor 
differences in atmospheric conditions, but provide a lower average detector noise 
background which is better suited to identify faint extended emission far from the V838 Mon stellar core.  
Figure \ref{psffig} illustrates 
our V838 Mon imagery following subtraction of optimally scaled and registered PSF-star observations, plotted 
on a linear stretch in units of mJy, and smoothed by a second-order Gaussian function.  We have masked the 
region inside a radial distance of 0$\farcs$8  (8-pixels), which is dominated by PSF-subtraction residuals, along 
with a diagonal strip in our N$^{'}$ imagery which exhibited a commonly observed detector artifact which arises when 
Michelle images bright sources.  

We detect no convincing evidence of extended emission above a level of $\sim$1 mJy at 11.2 $\mu$m 
and above a level of $\sim$7 mJy at 18.1 $\mu$m.  Applying additional binning 
or alternate smoothing functions to the data presented in Figure \ref{psffig} yielded no evidence of extended 
emission above the observed residual detector noise background.  

As described in Section \ref{phot}, we shifted and scaled the archival 2004 Spitzer MIPS 24$\mu$m data to extract 
point source photometry for the 2007 Spitzer 24$\mu$m data.  The scaling factor we applied to the 2004 data 
(1.97) led to an optimal cancellation of the PSF wings 
of the 2007 data in the subtraction process, but produced a large region of over-subtracted flux exterior to the 
stellar core, whose morphology corresponded to the extended emission reported by \citet{ban06}.  Using a 
circular aperture of radius 80$\farcs$0 and masking the central PSF-core region, we determined the net 
24$\mu$m flux density of this oversubtraction region was -0.88 Jy.  This is exactly the flux density one would 
expect to see if the extended emission component did not vary between 2004 and 2007, 
given the reported 2004 extended emission flux of 0.91 Jy \citep{ban06} and our use of a scaling factor of 1.97, i.e.
\begin{displaymath}
24\mu m_{2007}  - (1.97 \times 24\mu m_{2004}) = 24\mu m_{net}
\end{displaymath} 
\begin{displaymath}
0.91 Jy - (1.97 \times 0.91 Jy) = -0.88 Jy
\end{displaymath} 
While a comprehensive comparative analysis of the 2004 and 2007 Spitzer MIPS data is outside of 
the scope of this Letter, our simple analysis suggests a) the amount of 24$\mu$m and 70$\mu$m emission in 
the unresolved V838 Mon stellar core has increased by a factor of $\sim$2 from 2004 to 2007; and b) the 
net amount of 24$\mu$m extended emission appears unchanged from 2004 to 2007.

\section{Discussion} \label{disc}

We detect no 
evidence of extended 18.1 $\mu$m emission above a level of $\sim$7 mJy from radial distances of 
1860 AU (0$\farcs$3) to 93000 AU (15$\farcs$0), assuming a distance of 6.2 kpc \citep{spa08}.  Making 
the simple assumption that V838 Mon's circumstellar ejecta expands at a constant velocity, this gap of thermal IR 
emission suggests no circumstellar dust producing events have occurred within the past 
$\sim$900 (if v$_{ejecta}$ $\sim$300 km s$^{-1}$; \citealt{wi03a}) to $\sim$1500 years (if 
v$_{ejecta}$ $\sim$500 km s$^{-1}$; \citealt{mun02}).  Our analysis of 2007 imagery of V838 Mon from 
both Gemini's Michelle and Spitzer's MIPS 
indicates the presence of significantly enhanced 10-70$\mu$m emission in the unresolved central stellar core
in 2007 as compared to 2004.  We interpret our 2007 data as evidence that a \textit{new} 
circumstellar dust production event has occurred since 2004.  

Spectroscopic evidence of V838 Mon's binary companion disappeared and H$\alpha$ returned to emission 
in late 2006, corresponding with a temporary multi-color photometric fading \citep{bon06,mu07b}.  
\citet{mu07b} attributed 
these events to either a) an eclipse between the primary and secondary; or b) a dust cloud characterized by 
E$_{B-V}$ = 0.55 and R$_{V}$ = 3.1, gravitationally  
bound to the binary system, which eclipsed the B3V secondary.  We suggest alternate interpretations are possible 
if these photometric and spectroscopic events are correlated to the new mid-IR behavior we have 
found in data from early 2007.  A gravitationally bound dust cloud would radiate a constant amount of thermal 
IR emission throughout its orbit except during the brief time that it is occulted by the secondary; thus, \textit{unless 
it was only recently formed}, such a cloud could not be singularly responsible for producing the observed enhanced 
mid-IR emission.  A simple eclipse between the primary and secondary also 
would not lead to enhanced thermal IR emission.  

\citet{bon06} suggested that the photometric and 
spectroscopic events of late 2006 were indicative of the 2002 ejecta interacting with and overtaking the B3V 
binary.  More recent observations indicate that V838 Mon is experiencing a prolonged 
period of reduced photometric brightness (H. Bond 2008, personal communication), further supporting this 
interpretation.  While the temporal resolution of our data do not allow us to quantitatively determine 
whether this proposed encounter is related to the new circumstellar dust creation event we report, we suggest 
that such a correlation is plausible.  

We suggest that the expanding ejecta from the 2002 outbursts has condensed to form \textit{new} circumstellar dust, 
producing the enhanced IR excess observed inside a radial distance of 1860 AU.  The dust condensation radius 
of $\sim$3.5 AU for the B3V companion, which assumes T$_{cond}$ $\sim$1000K, $R_{cond} \sim\frac{R_{star}}{2} (\frac{T_{star}}{T_{cond}})^{2} $ \citep{ire05}, and the estimated primary-secondary separation of $\sim$34 AU 
derived from interferometry \citep{lan05}, indicates that dust can survive within the radial distance constraint 
of $<$1860 AU implied by our observations.  This type of event is not altogether unexpected, as \citet{lyn04} previously 
reported the detection of an expanding region of molecular gas which they noted was likely dust precursor  
material.  The enhanced flux at 18.1, 24, and 70 $\mu$m we observed in 2007 
suggests that the equally strong (2007 epoch) 11.2 $\mu$m 
flux might not represent an additional enhancement of gas-phase molecular emission above that suggested by 
\citet{lyn04}, but rather represents a warm thermal dust emission component.  

Early post-outburst polarimetric observations \citep{wi03a,des04} indicated this ejecta was 
asymmetric; hence, we 
suggest that it is likely that dust which condenses from this ejecta will be spatially non-uniform and/or 
clumpy.  We speculate 
that it is possible that one or more of these clumps eclipsed the line of sight in late 2006, producing the attenuation 
event reported by \citet{bon06} and \citet{mu07b}.  Unlike the dust cloud eclipse event invoked by \citet{mu07b}, 
we believe that any eclipsing dust blob is more likely to located in the expanding ejecta which is condensing to form 
new dust.  Followup interferometric observations might reveal differences in the angular size of the system 
from that found in 2004 epoch observations by \citet{lan05} if this ejecta has indeed recently reached the location 
of the B3V secondary star.

\acknowledgments

We thank H. Bond, B. Bonev, M. Perrin, A. Roberge, and K. Volk for helpful 
discussions about our data, and our annonymous referee for providing 
feedback which helped to improve this paper.  Support for this project was provided by a 
NASA NPP fellowship to JPW (NNH06CC03B).

\clearpage
\newpage
\begin{table} 
\begin{center}
\footnotesize
\caption{Summary of Observations and Flux Densities\label{obstable}}
\begin{tabular}{lccccccc}
Target & UT Date & Observatory & $\lambda_{central}$ & $\delta\lambda$ & FWHM &  Flux Density & Comment \\
 & & & ($\mu$m) & ($\mu$m) & (arcsec) & (mJy) &  \\
\tableline
V838 Mon & 2004 October 14$^{1}$ & Spitzer & 23.7 & 4.7 & 6$\farcs$0 & 15.06$\pm$0.75$^{2}$ & \citet{ban06}  \\
V838 Mon & 2005 April 1$^{1}$ & Spitzer & 71.42 & 19 & 18$\farcs$0 & 3.82$\pm$0.38$^{2}$ & \citet{ban06}  \\
HD 52666 & 2007 March 21 & Gemini & 11.2 & 2.4 & 0$\farcs$40 & \nodata & PSF-star obs \\
HD 66141 & 2007 March 21 & Gemini & 11.2 & 2.4 & 0$\farcs$43 & \nodata & flux std \\
V838 Mon & 2007 March 21 & Gemini & 11.2 & 2.4 & 0$\farcs$40 & 29.54$\pm$0.14$^{3}$ & \nodata   \\
V838 Mon & 2007 March 22 & Gemini & 18.1 & 1.9 & 0$\farcs$61 & 36.43$\pm$0.82$^{3}$ & \nodata  \\
HD 52666 & 2007 March 22 & Gemini & 18.1 & 1.9  & 0$\farcs$57 & \nodata &  PSF-star \\
HD 60522 & 2007 March 22 & Gemini & 18.1 & 1.9 & 0$\farcs$59 & \nodata & flux std  \\
V838 Mon & 2007 April 10 & Spitzer & 23.7 & 4.7  & 6$\farcs$0 & 29.67$\pm$$^{1.95}_{1.91}$$^{2}$ & AOR-21475072 \\
V838 Mon & 2007 April 10 & Spitzer & 71.42 & 19 & 18$\farcs$0 & 7.34$\pm$0.73$^{2}$ & AOR-21475072  \\
HD 168723 & 2007 August 6 & Gemini & 18.1 & 1.9 & 0$\farcs$44 & \nodata & deep PSF-star obs \\
HD 168723 & 2007 August 9 & Gemini & 11.2 & 2.4 & 0$\farcs$54 & \nodata & deep PSF-star obs  \\
\tableline
\tablecomments{$^{1}$ 2004/2005 Spitzer flux density measurements were adopted from Table 2 of \citet{ban06}.  
$^{2}$ The uncertainties quoted for Spitzer flux density measurements, 5\% for 24$\mu$m and 10\% for 70$\mu$m, 
represent the estimated absolute flux uncertainty.  $^{3}$  Gemini flux density error estimates represent the internal 
uncertainties of the measurements; additional absolute flux scaling uncertainties are expected to be of order several percent for bright sources observed in stable conditions (K Volk 2008, 
personal communication).}
\end{tabular}
\end{center}
\end{table}

\clearpage
\newpage

\clearpage
\newpage
\begin{figure}
\includegraphics[scale=0.6]{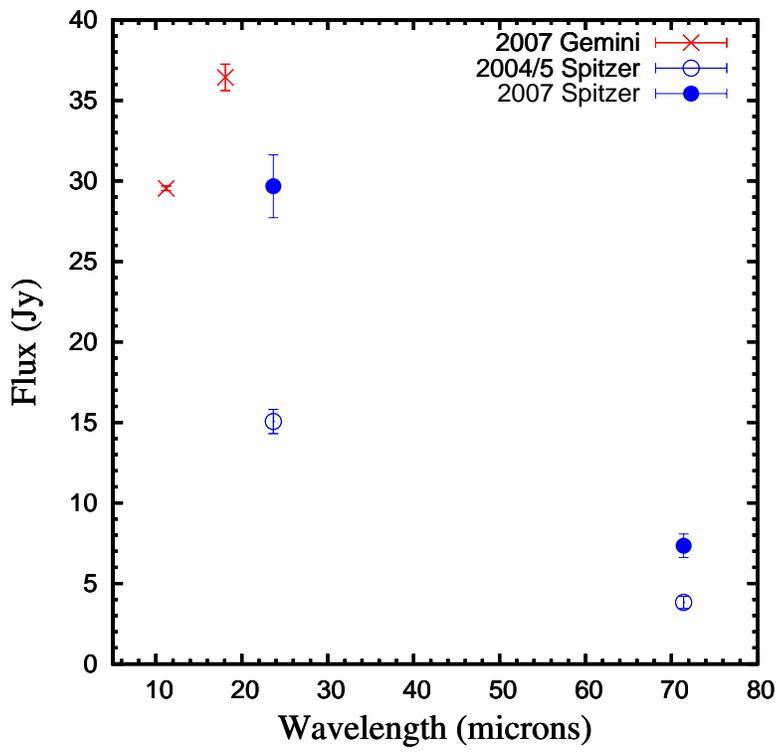} 
\figcaption[f2.eps]{SED of the unresolved point source flux of V838 Mon, based on the flux densities listed 
in Table \ref{obstable}.  The 2007 Gemini and Spitzer data, obtained $\sim$2 weeks apart, are consistent 
to within 3$\sigma$ of the total error budget of each observation.  The IR excess emission at $\sim$20$\mu$m 
and 70$\mu$m has clearly increased between 2004 and 2007.  We interpret this as evidence that a 
circumstellar dust production event has recently occurred. \label{sedfig}}
\end{figure}

\clearpage
\newpage


\begin{thebibliography}{}

\bibitem[Afsar \& Bond(2007)]{afs07} Afsar, M. \& Bond, H.E. 2007, AJ, 133, 387
\bibitem[Banerjee et al.(2006)]{ban06} Banerjee, D.P.K., Su, K.Y.L., Misselt, K.A., \& Ashok, N.M. 2006, ApJL, 644, 57
\bibitem[Bond et al.(2003)]{bon03} Bond, H.E. 2003, Nature, 422, 405
\bibitem[Bond(2006)]{bon06} Bond, H.E. 2006, ATel, 966 
\bibitem[Brown et al.(2002)]{bro02} Brown, N.J. et al. 2002, IAU Circ. 7785
\bibitem[Cohen et al.(1999)]{coh99} Cohen, M., Walker, R.G., Carter, B., Hammersley, P., Kidger, M., 
\& Noguchi, K. 1999, AJ, 117, 1864
\bibitem[Desidera \& Munari(2002)]{des02} Desidera, S. \& Munari, U. 2002, IAU Circ. 7982
\bibitem[Desidera et al.(2004)]{des04} Desidera, S. et al. 2004, A\&A, 414, 591
\bibitem[Evans et al.(2003)]{eva03} Evans, A., Geballe, T.R., Rushton, M.T., Smalley, B., van Loon, J. Th., 
Eyres, S.P.S., \& Thne, V.H. 2003, MNRAS, 343, 1054
\bibitem[Henden et al.(2002)]{hen02} Henden, A., Munari, U., \& Schwartz, M. 2002, IAU Circ. 7859
\bibitem[Ireland et al.(2005)]{ire05} Ireland, M.J., Tuthill, P.G., Davis, J., \& Tango, W. 2005, MNRAS, 361, 337
\bibitem[Kimeswenger et al.(2002)]{kim02} Kimeswenger, S., Lederie, C., Schmeja, S., \& Armsdorfer, B. 
2002, MNRAS, 336, L43
\bibitem[Lanet et al.(2005)]{lan05} Lane, B.F., Retter, A., Thompson, R.R., \& Eisner, J.A. 2005, ApJL, 622, 137
\bibitem[Lawlor(2005)]{law05} Lawlor, T.M. 2005, MNRAS, 361, 695
\bibitem[Lynch et al.(2004)]{lyn04} Lynch, D.K. et al. 2004, ApJ, 607, 460
\bibitem[Munari et al.(2002)]{mun02} Munari, U. et al. 2002, A\&A, 389, 51
\bibitem[Munari et al.(2005)]{mun05} Munari, U. et al. 2005, A\&A, 434, 1107
\bibitem[Munari et al.(2007a)]{mun07} Munari, U., Navasardyan, H., \& Villanova, S. 2007a, in The Nature of V838 Mon and Its Light Echo, ed. R.L.M. Corradi \& U. Munari, ASP Conf. Ser., 363, 13
\bibitem[Munari et al.(2007b)]{mu07b} Munari, U. et al. 2007, A\&A, 474, 585
\bibitem[Rauch et al.(2002)]{rau02} Rauch, T., Kerber, F., \& Van Wyk, F. 2002, IAU Circ. 7886
\bibitem[Retter et al.(2006)]{ret06} Retter, A., Zhang, B., Siess, L., \& Levinson, A. 2006, MNRAS, 370, 1573
\bibitem[Rieke et al.(2004)]{rie04} Rieke, G.H. et al. 2004, ApJS, 154, 25
\bibitem[Rushton et al.(2005)]{rus05} Rushton, M.T. et al. 2005, MNRAS, 360, 1281
\bibitem[Sparks et al.(2008)]{spa08} Sparks, W.B. 2008, AJ, 135, 605
\bibitem[Tylenda et al.(2005)]{tyl05} Tylenda, R., Soker, N., \& Szczerba, R. 2005, A\&A, 441, 1099
\bibitem[Tylenda \& Soker(2006)]{tyl06} Tylenda, R. \& Soker, N. 2006, A\&A, 451, 223 
\bibitem[Wisniewski et al.(2003a)]{wi03a} Wisniewski, J.P. et al. 2003a, ApJ, 588, 486
\bibitem[Wisniewski et al.(2003b)]{wi03b} Wisniewski, J.P., Bjorkman, K.S., \& Magalhaes, A.M. 2003b, ApJL, 598, 43
\bibitem[Wisniewski(2007)]{wis07} Wisniewski, J.P. 2007, in The Nature of V838 Mon and Its Light Echo, ed. R.L.M. 
Corradi \& U. Munari, ASP Conf. Ser., 363, 73

\end{thebibliography}
\end{document}